\documentclass[acmtog, anonymous=false, review=false]{acmart}

\usepackage{booktabs} 

\setcitestyle{square}

\usepackage[mathscr]{euscript}
\usepackage{amsmath}
\usepackage{mathtools}
\usepackage{stackrel}
\usepackage{nicefrac}

\usepackage{algpseudocode}
\usepackage[ruled]{algorithm2e} 

\SetAlFnt{\small}
\SetAlCapFnt{\small}
\SetAlCapNameFnt{\small}
\SetAlCapHSkip{0pt}
\IncMargin{-\parindent}

\usepackage{dblfloatfix}
\usepackage{float}

\usepackage{siunitx}
\usepackage{subcaption}

\acmJournal{TOG}
\acmVolume{0}
\acmNumber{0}
\acmArticle{0}
\acmYear{2022}
\acmMonth{0}

\begin{document}

\title{Pupil-aware Holography} 

\author{Praneeth Chakravarthula}
\affiliation{%
	\institution{Princeton University}
}

\author{Seung-Hwan Baek}
\affiliation{%
    \institution{POSTECH and Princeton University}
}

\author{Florian Schiffers}
\affiliation{%
	\institution{Northwestern University}
}

\author{Ethan Tseng}
\affiliation{%
	\institution{Princeton University}
}

\author{Grace Kuo}
\author{Andrew Maimone}
\author{Nathan Matsuda}
\affiliation{%
	\institution{Meta Reality Labs}
}

\author{Oliver Cossairt}
\affiliation{%
	\institution{Northwestern University and Meta Reality Labs}
}

\author{Douglas Lanman}
\affiliation{%
	\institution{Meta Reality Labs}
}

\author{Felix Heide}
\affiliation{%
	\institution{Princeton University}
}

\authorsaddresses{}

\renewcommand\shortauthors{Chakravarthula et al.}

\begin{abstract}
Holographic displays promise to deliver unprecedented display capabilities in augmented reality applications, featuring a wide field of view, wide color gamut, spatial resolution, and depth cues all in a compact form factor. While emerging holographic display approaches have been successful in achieving large {\'e}tendue and high image quality as seen by a camera, the large {\'e}tendue also reveals a problem that makes existing displays impractical: the sampling of the holographic field by the eye pupil. Existing methods have not investigated this issue due to the lack of displays with large enough {\'e}tendue, and, as such, they suffer from severe artifacts with varying eye pupil size and location.

We show that the holographic field as sampled by the eye pupil is highly varying for existing display setups, and we propose pupil-aware holography that maximizes the perceptual image quality irrespective of the size, location, and orientation of the eye pupil in a near-eye holographic display. We validate the proposed approach both in simulations and on a prototype holographic display and show that our method eliminates severe artifacts and significantly outperforms existing approaches. 
\end{abstract}

\begin{teaserfigure}  
\vspace{3mm}
  \includegraphics[width=\linewidth]{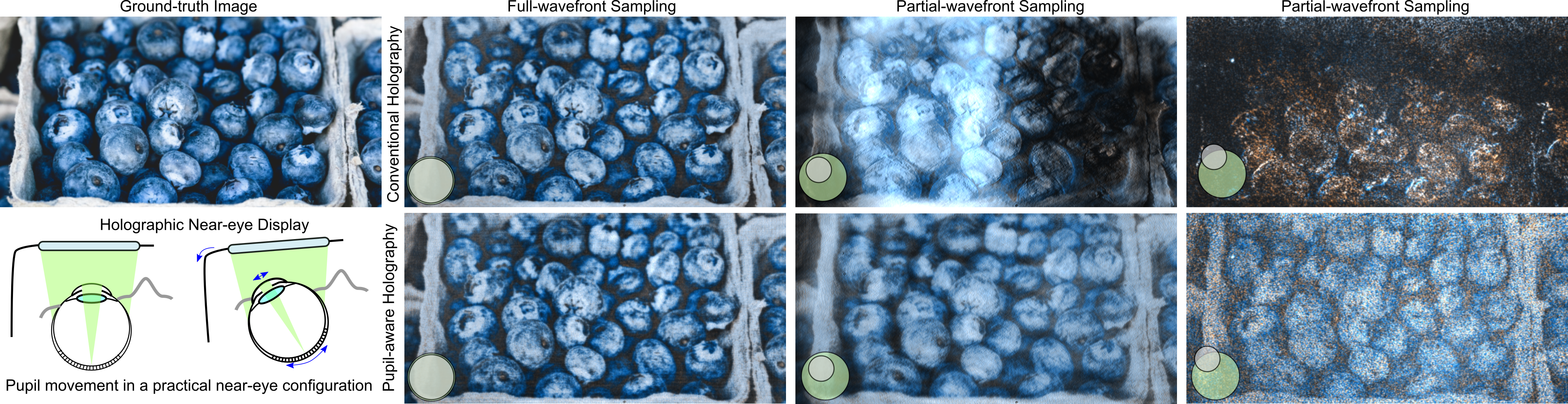}
  \caption{
  \emph{Pupil-aware holography on experimental hardware.} Conventional holographic displays often use a large lens to acquire the entire wavefront for high-fidelity reconstructions, but resulting in a tiny effective eyebox. However, near-eye holographic displays pose a unique problem in that the wavefront can only be partially sampled by the moving eye pupil with an unknown location and diameter at any given instant. This results in catastrophic failures such as complete loss of image on existing holographic displays (top-right). In this paper, we present the first pupil-aware near-eye holography framework that identifies and addresses this pupil-dependency problem, achieving robust reconstructions for arbitrarily partially sampled wavefronts, even at the edge of the eyebox (bottom-right). All of the results shown in this figure are acquired on an experimental hardware prototype.
  }
  \label{fig:teaser}
  \vspace{3mm}
\end{teaserfigure}

\maketitle


\newcommand{\vect}[1]{\mathbf{#1}}
\newcommand{\mat}[1]{\mathbf{#1}}

\newcommand\note[1]{\textcolor{red}{#1}}

\newcommand{\psf}{\rho}

\newcommand{\argmin}[1]{\stackrel[\{ #1 \}]{}{\textrm{arg min}}}
\newcommand{\minimize}[1]{\stackrel[\{ #1 \}]{}{\textrm{minimize}}}

\section{Introduction}
\label{sec:intro}
Augmented and virtual reality is emerging as a future technology with the potential to solve long standing challenges in human-computer interaction across domains, enabling applications as diverse as telepresence, surgical training, and automotive display. The key enabler of immersive augmented and virtual reality (AR/VR) are near-eye displays that are ultra compact and offer a wide field of view, high image quality and natural depth cues. Today, holographic displays are the only display technology that promises such unprecedented capabilities. 

In a holographic display, an input wave field is modulated typically by a phase-only spatial light modulator (SLM). The modulated wavefront then propagates to a distance to create the desired image as an interference pattern. This image formation stands in contrast with today's ray-based displays such as LCD or LED displays. The control over the entire wavefront of light enables holographic displays to create images with drastically reduced optical stacks by encoding much of the optics into the SLM phase pattern~\cite{Wakunami2016ProjectiontypeSH,maimone2020holographic}, thereby enabling the compact form factors and multi-focal capabilities required for near-eye display applications.

Recently, researchers have successfully demonstrated holographic displays that achieve image quality almost matching that of conventional displays~\cite{shi2021towards,chakravarthula2020learned,peng2020neural}. These methods lift restrictions on holographic phase computation by using neural network predictors and cameras in-the-loop to calibrate the display setups. Moreover, today's holographic displays are subject to limited {\'e}tendue, which means, that for 1K SLMs, one must heavily trade-off the eyebox (exit pupil) size for FoV. Emerging large {\'e}tendue displays are also starting to address this limitation, either naively with smaller SLM pixel pitch < $4 \mu$m, such as the Holoeye GAEA-2 4K SLM, or using {\'e}tendue-expanding diffractive elements~\cite{kuo2020high}. Although this new breed of displays holds the promise of practical high-quality holographic displays with a wide FOV and large eyebox simultaneously, and without the need for eye tracking or pupil steering, unfortunately, the large {\'e}tendue comes at the cost of a high degree of noise that is pupil-dependent with varying pupil position and size  -- this pupil variance, unobserved in low {\'e}tendue setups makes existing large {\'e}tendue displays impractical. Specifically, even with an ideal modulator at 1 billion pixels and perfect phase and amplitude modulation, existing holographic displays are fundamentally subject to the problem of pupil variance.

The human eye pupil swivels over a large area due to factors ranging from involuntary saccades to voluntary gaze changes of the viewer~\cite{bahill1975main}, mandating a wide eyebox. However, a large eyebox comes at the cost of image quality that is not uniform across the entire eyebox and drastically suffers with pupil sampling over the entire eyebox due to partial wavefront observations as shown in Fig~\ref{fig:teaser}. Depending on the size and orientation of the pupil, these perceptual artifacts can be widely different and quickly change. A Maxwellian-style display with a tiny eyebox in combination with eye tracking is one option to partly mitigate this problem~\cite{maimone2017holographic}. However, should the latency of the eye tracker do not match the eyebox movement, the user would perceive no image~\cite{jang2019holographic}. 
While recent large {\'e}tendue approaches, in theory, lift the need for eye tracking, the pupil variance in these emerging systems has not been investigated in the past as existing displays did not provide enough {\'e}tendue to observe this issue which is essential to make such displays practical. 

In this work, we investigate the \emph{unexplored problem of image quality dependence on eye pupil} for existing displays, and \emph{propose a holographic phase retrieval method} that is aware of the different pupil states, lifting the dependency of the image quality on the exit pupil in existing holography systems. The proposed system does not rely on eye-tracking but instead optimizes the phase pattern on the SLM to produce consistent projections independently of the pupil location and size. To investigate pupil variance in the presence of large {\'e}tendue, instead of adopting an {\'e}tendue expanding element~\shortcite{kuo2020high}, which mandates a complex calibration and alignment procedure, we use a system with limited {\'e}tendue to emulate a small part of the FOV of a large {\'e}tendue system (with large FOV and eyebox). This allows us to prototype the proposed method with affordable SLMs using a larger eyebox and smaller FOV system, and investigate pupil effects across the eyebox. Note that simulating a large {\'e}tendue display on a low {\'e}tendue display is a valid experiment as it just has a smaller FOV. However, the proposed algorithmic approach only considers the image fidelity across the eyebox and hence, irrespective of the FOV, holds for future large {\'e}tendue displays as well. 

Being able to study various pupil dependent artifacts at several pupil states, we propose a display algorithm that generates high quality images across the eyebox regardless of the eye pupil size and location. We introduce a differentiable pupil-aware image formation model and a corresponding optimization method that incorporate sampling over diverse pupil states, that, together, ensure pupil awareness when solving for SLM patterns that result in high quality reconstructions across the eyebox.

Pupil-aware holography brings an under-investigated problem of holographic displays to attention and proposes a method to overcome the problem of dynamic pupil sampling of the eyebox. The proposed approach takes a step forward to enable practical holographic displays for immersive augmented and virtual reality. 

In particular, we make the following contributions
\begin{itemize}
    \setlength\itemsep{0.4em}
    \item We \emph{analyze} the dependence of image quality on the eye pupil states and energy distribution across the eyebox for various holographic displays.
	\item We \emph{introduce} a new differentiable pupil-aware holographic image formation method that accounts for different pupil positions and orientations within the eyebox.
	\item We devise a \emph{content-aware phase optimization} method that allows us to learn optimal hologram phase patterns which maintain intensity across the eyebox without introducing image artifacts and severe speckle.
	\item We \emph{analyze} our approach in simulation and demonstrate significant higher image quality across the eyebox compared to existing holographic display methods. Using a prototype system, we \emph{validate experimentally} that the proposed pupil-aware holography approach achieves significant reduction in pupil-dependent artifacts.

\end{itemize}

\begin{figure}[t]
    \centering
    \includegraphics[width=0.98\linewidth]{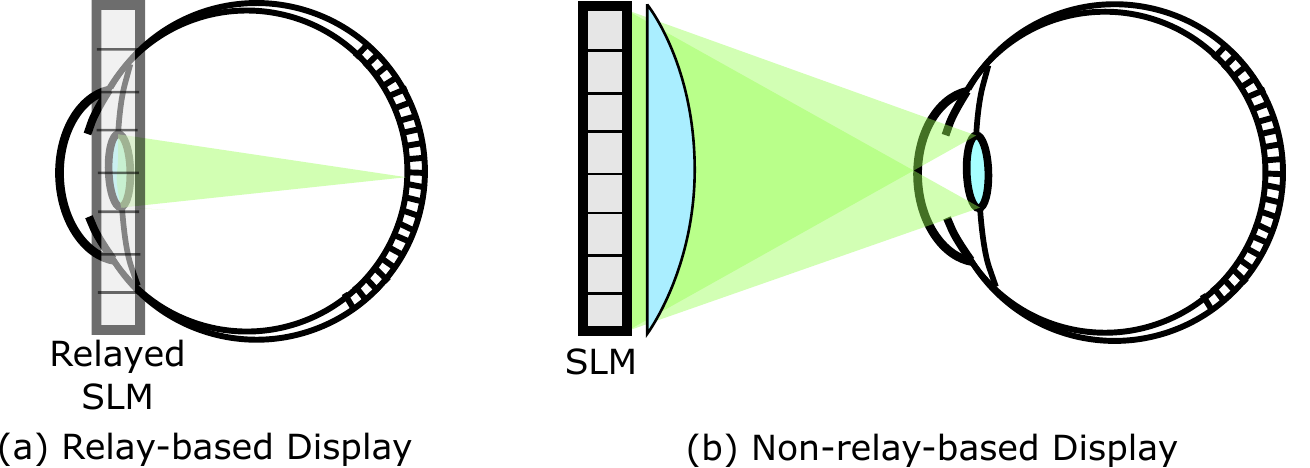}
    \caption{
    Existing holographic near-eye display designs either (a) directly sample diffracted light from the SLM by relaying the virtual SLM to the pupil plane (left), or (b) by projecting the diffracted light to a narrow valid region on the pupil plane using a lens (right). These designs, combined with a computer-generated hologram (CGH) algorithm form high-quality visual images if the entire wavefront is fully sampled by the eye pupil. Unfortunately, the eye pupil position, orientation and size dynamically change, resulting in partial and incomplete sampling of the diffracted light wavefront, causing severe artifacts in all existing holographic near-eye display systems.
    }
    \label{fig:baseline_display_designs}
\end{figure}

\subsection{Scope and Limitations}
This paper does not contribute new hardware but the analysis of an unexplored problem in holography and an algorithm to address it irrespective of the setup. For the first time, we achieve robust image fidelity and eyebox energy across the eyebox. To study pupil variance, we use a large eyebox display as a test bed, albeit with limited field of view. Irrespective of the FoV, the current approach only considers the image fidelity across the eyebox and hence holds for future large etendue displays with smaller SLM pixel pitch or larger area. We validate the applicability of the proposed method to future large etendue systems in simulation.

Human eyes distort the phase of incident light waves, deviating from an ideal thin lens, given the conditions of each human observer~\cite{chakravarthula2021gaze}, which we do not consider in this work. We do also not incorporate the holistic perception of the human visual system into the proposed method but only make a first step by addressing the display with varying pupil states. Incorporating further aspects of the perception of human visual system into our method may prove as exciting future work.

\section{Related Work}
\label{sec:related}

\begin{figure}
    \centering
    \includegraphics[width=\columnwidth]{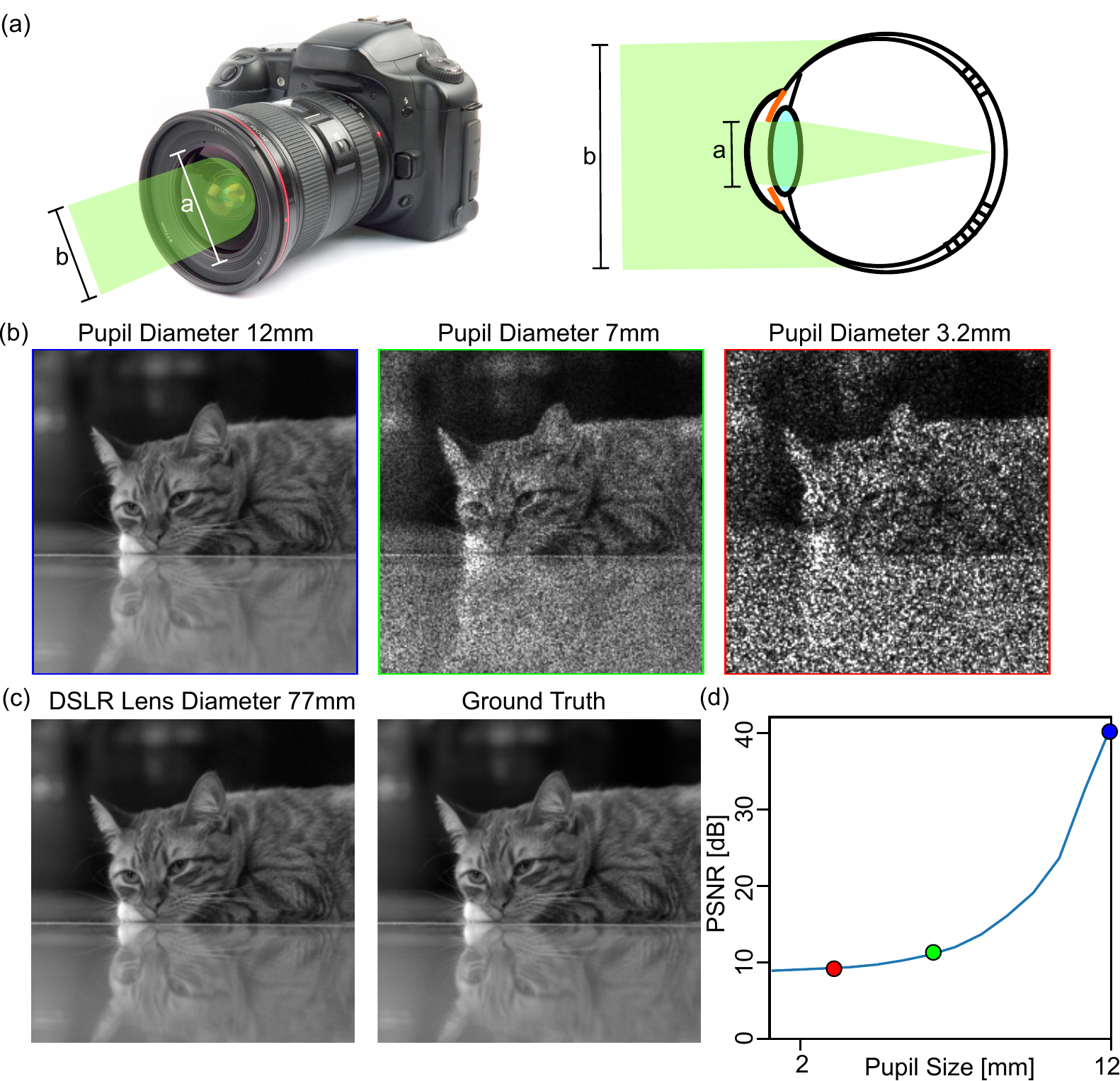}
    \caption{\label{fig:motivation}%
    (a) Compound DSLR lenses with large aperture $a$ are capable of capturing the entire light bundle with diameter $b$ emitted from a holographic display, as opposed to a human eye. (c) This results in high-fidelity holographic reconstructions on the imaging sensor. However, human eyes have limited- and variable-size pupils that only partially sample the incoming wavefront to the eyes. (b) \& (d) As a result of the partial wavefront sampling due to an unknown pupil state, the projected image quality degrades, as a conventional computer-generated holography assumes that all of the wavefront is properly captured. Note that the low-contrast region of the cat's reflection on the floor is close to being invisible for smaller pupil diameters.
    }
\end{figure}

\subsection{Holographic Displays}
Holographic displays promise to be the ultimate virtual and augmented reality displays of the future, potentially achieving high projection quality with depth cues in a thin eye-glasses form factor~\cite{maimone2017holographic, peng2020neural, chakravarthula2020learned}. The most successful approaches have achieved the highest image quality only with large benchtop prototype displays~\cite{shi2021towards}. At the same time, researchers have shown that a combination of holographic optical elements and dynamic holography can achieve sunglasses-like form factor for both AR and VR applications~\cite{maimone2017holographic}. All of these methods have in common that they are constrained by the low \'etendue of existing SLMs, limiting either the FoV or eyebox of the displays. To address this issue, researchers have proposed methods for more effective use of \'etendue of holographic displays by incorporating eyetracking and dynamically moving the small eyebox or statically expanding the eyebox. Such an eyebox expansion can be achieved by pupil steering where the incident illumination on the SLM is changed in demand~\cite{jang2019holographic} or by static eyebox expansion where several copies of the exit pupil are created~\cite{xia2020towards}. Unfortunately, these pupil steering methods mandate precise and low-latency eyetracking, along with complex and bulky optics. Without eyetracking present, no existing methods have considered the non-uniform sampling of eye pupil and the complex artifacts that arise across the eyebox. In this work, we analyze pupil variance in existing work and we propose a pupil-aware hologram generation approach for wide eyebox holographic displays. 

\subsection{\'Etendue Expansion for Holographic Displays}
Existing approaches for the expansion of \'etendue in holographic displays employ a diffractive mask with higher pixel pitch than the SLM employed~\cite{buckley2006viewing,park2019non,kuo2020high} to expand the largest diffraction angle that existing SLM devices can support. However, these existing methods were limited to generating sparse projections such as a few foci or letters. Buckley et al.~\shortcite{buckley2006viewing} used a diffractive phase mask in front of a binary phase SLM to create a static display where the twin images were removed and the viewing angle was increased simultaneously. Park et al.~\shortcite{park2019non}, on the other hand, used a known diffractive amplitude mask to achieve significant increase in {\'e}tendue, but only could generate sparse focal spots simultaneously. Recently, Kuo et al.~\shortcite{kuo2020high} extended this idea to generate dense, photo-realistic \'etendue expanded holograms at the native resolution of the SLM. To make for an efficient testbed for the proposed method, instead of increasing {\'etendue}, which is an active direction that today requires manufacturing, alignment and calibration of diffractive elements, we use a focusing lens in front of the SLM to achieve a large eyebox. While similar hardware configurations have been proposed in the past~\cite{maimone2017holographic} to achieve a large FOV, we propose a variant with large eyebox, which allows us to investigate pupil variance and validate the proposed method when only observing a partial eyebox.

\subsection{Computer Generated Holography Algorithms}
\paragraph{Traditional Phase Retrieval Methods}
Hologram generation algorithms find the appropriate SLM pattern for a given target image displayed by the system. This is a particularly challenging task with existing phase-only SLM devices available today. Holographic phase retrieval approaches can be broadly classified into single-step methods and iterative methods. While single-step methods such as amplitude-discard or double-phase amplitude coding~\cite{maimone2017holographic} may provide sufficient image quality for some applications, iterative algorithms such as the popular Gerchberg-Saxton method~\cite{gerchberg1972practical} significantly improve the image fidelity. Some of the early iterative methods for holographic phase retrieval include error reduction using iterative optimization~\cite{lesem1969kinoform,gerchberg1972practical}, together with an assumption on a non-zero support of the real-valued signal. One of many extensions of such iterative algorithm is the popular hybrid input-output (HIO) method~\cite{fienup1982phase}, and others with various relaxations~\cite{luke2004relaxed,bauschke2003hybrid}. Researchers have explored using alternating direction methods for phase retrieval~\cite{wen2012alternating,marchesini2016alternating}, non-convex optimization~\cite{zhang20173d} and overcoming the non-convex nature of the phase retrieval problem by lifting, i.e., relaxation, to a semidefinite~\cite{candes2013phaselift} or linear~\cite{goldstein2018phasemax,bahmaniPhaseMax} program. Recent iterative phase retrieval methods using first-order stochastic gradient descent with complex Wirtinger gradients~\cite{chakravarthula2019wirtinger,peng2020neural} have been able to achieve high image quality on holographic displays. Such iterative optimization approaches were recently used to produced high-quality 3D holograms~\cite{choi2021neural,chakravarthula2022hogel}.

\paragraph{Neural Phase Retrieval Methods}
Neural networks and deep learning approaches have recently been proposed as tools for optical design and holographic phase retrieval. Researchers have tackled holographic microscopy by solving phase retrieval problems using neural networks~\cite{rivenson2018phase,eybposh2020deepcgh}. In a similar fashion, neural networks have been investigated for learning holographic wave propagation from a large training dataset. For example, Horisaki et al.~\shortcite{horisaki2018deep} trained a U-net on a pair of SLM phase and intensity patterns, and predicted SLM phase patterns during inference. Recently, Eybposh et al.~\shortcite{eybposh2020deepcgh} proposed an unsupervised training strategy and predicted the SLM phase patterns in real-time that produced 2D and 3D holographic projections. Peng et al.~\shortcite{peng2020neural} and Chakravarthula et al.~\shortcite{chakravarthula2020learned} have recently demonstrated camera-in-the-loop (CITL) calibration of hardware using neural networks and high fidelity holographic images on prototype displays. Shi et al.~\shortcite{shi2021towards} have demonstrated a light-weight phase retrieval network and Wang et al.~\shortcite{wang2022dprc} further showed neural hologram compression framework that may be suitable for inference on mobile hardware in the future.

The proposed pupil-aware holography approach shares with prior work that we solve an optimization problem to find optimal SLM patterns for a target image. However, we show that existing method, although achieving high image quality, are not practical due to pupil variance. We propose a pupil-aware holographic display approach where we formulate the loss function to include several pupil states within a large eyebox and compensate for perceptual artifacts across the eyebox.

\section{Pupils in Near-eye Holographic Displays}
\label{sec:fwdmodels}

As a result of the limited bandwidth and pixel count of SLMs in the past, most existing approaches in near-eye holographic displays had to maximize the FOV at the cost of a small eyebox. Large FOV together with large eyebox has recently been achieved by using \'{e}tendue expanding methods~\cite{kuo2020high}. We describe here the complications that arise with pupil variance and thereby partial wavefront sampling for all of these existing display types. To this end, we consider pupil variance for holographic displays in two possible configurations where: (1) the SLM is relayed directly onto the eye pupil (relay-based displays), and (2) the SLM is away from the eye (non-relay-based displays). 

\begin{figure}[t]
    \centering
    \includegraphics[width=0.9\columnwidth]{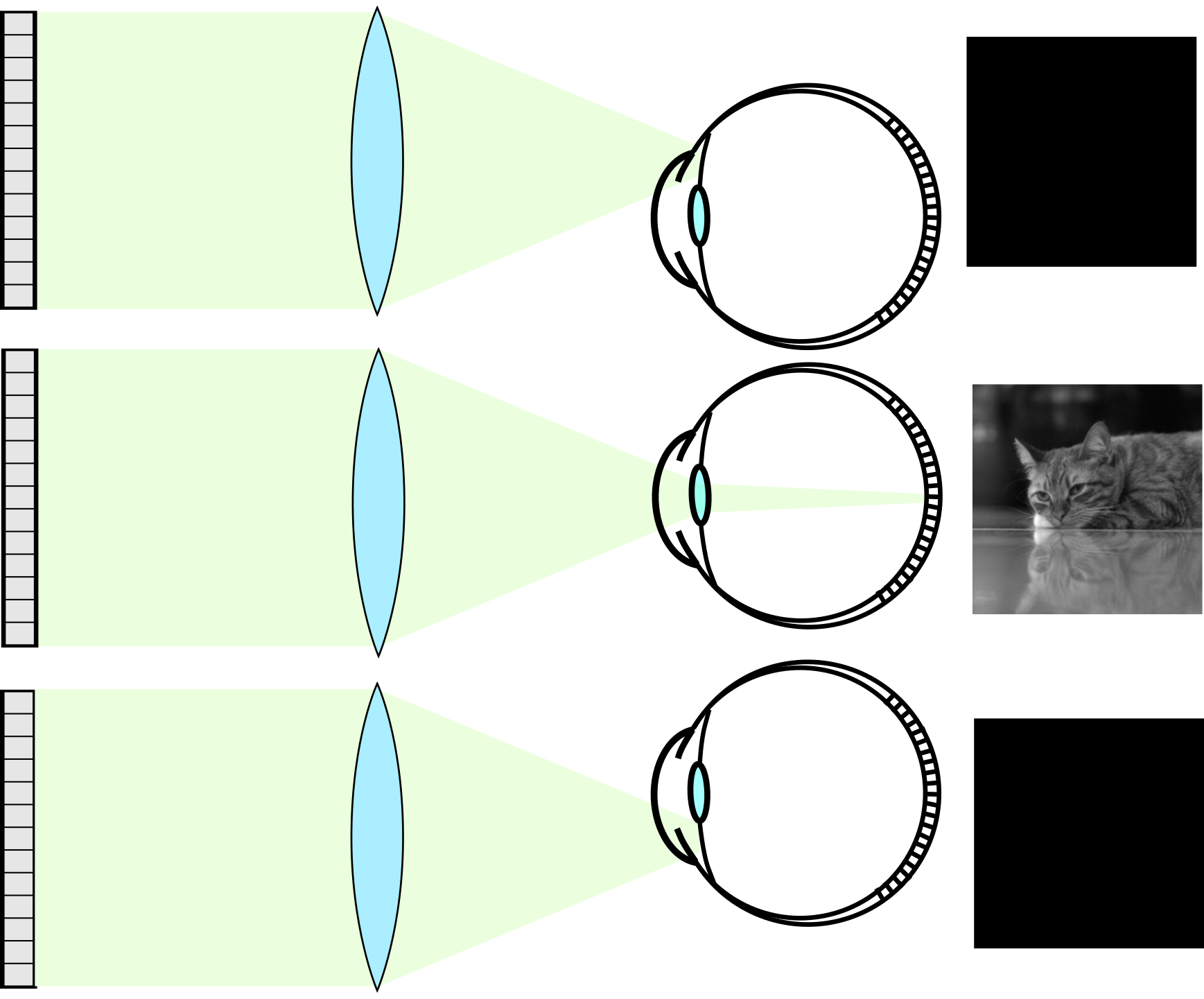}
    \caption{
    Conventional holographic displays~\cite{maimone2017holographic,kim2022holographicglasses} focus the SLM modulated light down to a point on the pupil plane, resulting in a very tiny eyebox (middle). This approach severely restricts the ``valid'' region where the eyes can see the holographic image, thereby resulting in no viewable imagery outside of the small eyebox region (top and bottom) that is mandated by the limited {\'etendue} of today's holographic systems.
    }
    \label{fig:nonrelayed}
\end{figure}

\paragraph{Pupils in Relay-based Displays}
In holographic displays with a relay, as illustrated in Fig.~\ref{fig:baseline_display_designs}(a), the wavefront at the SLM is relayed onto (or sometimes close to) the eye pupil plane and focused onto the retina. In this configuration, near-field or far-field holograms can be generated by controlling the $z$ distance of the target holographic image from the relayed SLM. If the complex wavefront at the SLM is $H$, the observed image intensity after the eye pupil sampling is given by
\begin{equation}
    I = | \mathcal{P}(H \odot M) |^2 ,
\end{equation}
where $\mathcal{P}$ is a wave propagation operator, $\odot$ is element-wise Hadamard product and $M$ is an amplitude mask, modeling the pupil sampling of the wavefront at the SLM plane, by the eye pupil.
The position and shape of the mask depends on the pupil location, size and orientation.
The propagation operator $\mathcal{P}$ is a Fresnel propagation function for a near-field hologram and a Fraunhofer or Fourier propagation function for a far-field hologram, respectively.

As the propagating wavefront is apodized, a significant portion of the light from the SLM does not enter the eye pupil. A far-field holographic image mimics light coming from optical infinity and hence the pupil sampling generally results in reduced intensity in the reconstructed images. However, masking a near-field holographic wavefront results in complete loss of information from masked wavefront and hence the reconstructed images appear cropped, see Figure~\ref{fig:simulation_results} and Figure~\ref{fig:real_results}.
Not only does this mean that pupil sampling comes at the expense of SLM bandwidth, it also results in under-sampling of spatial frequencies (or the angular rays) from the SLM. This produces poor perceived quality in both near- and far-field configurations. 
Specifically, in the far-field configuration, the small pupil sizes result in extreme speckle artifacts due to the missing frequencies outside of the sampled wavefront as shown in Figure~\ref{fig:motivation} and Figure~\ref{fig:simulation_results}. 
Note that each pixel on the retina receives partial wavefront sampled by the pupil.
For the near-field configuration, partial sampling of the wavefront by a pupil state causes both cropping and diffraction artifacts from the wave propagation as shown in Figure~\ref{fig:simulation_results}.

\paragraph{Pupils in Non-relay-based Displays} 

In order to avoid undersampling of the wavefront by the pupil, non-relay-based displays use additional optics such as an eyepiece to converge the SLM modulated light into the eye pupil, instead of relaying the virtual SLM onto the eye.
Note that all SLM pixels are visible in the eyebox in this display configuration, and therefore the image formation model can be formalized by
\begin{equation}
    I = | \mathcal{P}(H) |^2 ,
\end{equation}
where $\mathcal{P}$ is a wave propagation function.
While all SLM pixels are exploited in this configuration due to the {\'e}tendue conservation, it comes at the cost of the eyebox being shrunk into a tiny viewable region around the focal spot of the eyepiece optics. The user would only see the hologram if their pupil lies directly on the focal spot. However, once the pupil moves away, the user would cease to view the image entirely, as depicted in Figure~\ref{fig:nonrelayed}.
Consequently, the eyebox needs to be expanded or replicated using additional optical elements~\cite{jang2017retinal} or steered in accordance with the pupil movement~\cite{kim2022holographicglasses} for such displays. 
\\

Overall, all the discussed \emph{existing display configurations do not adequately account for eye pupil movement and varying pupil size}, even if the eyebox is sufficiently large. We propose a method in the next section that addresses these limitations.

\section{Pupil-aware Holography}
\label{sec:pupil-invariant}
\begin{figure}
    \centering
    \includegraphics[width=0.98\columnwidth]{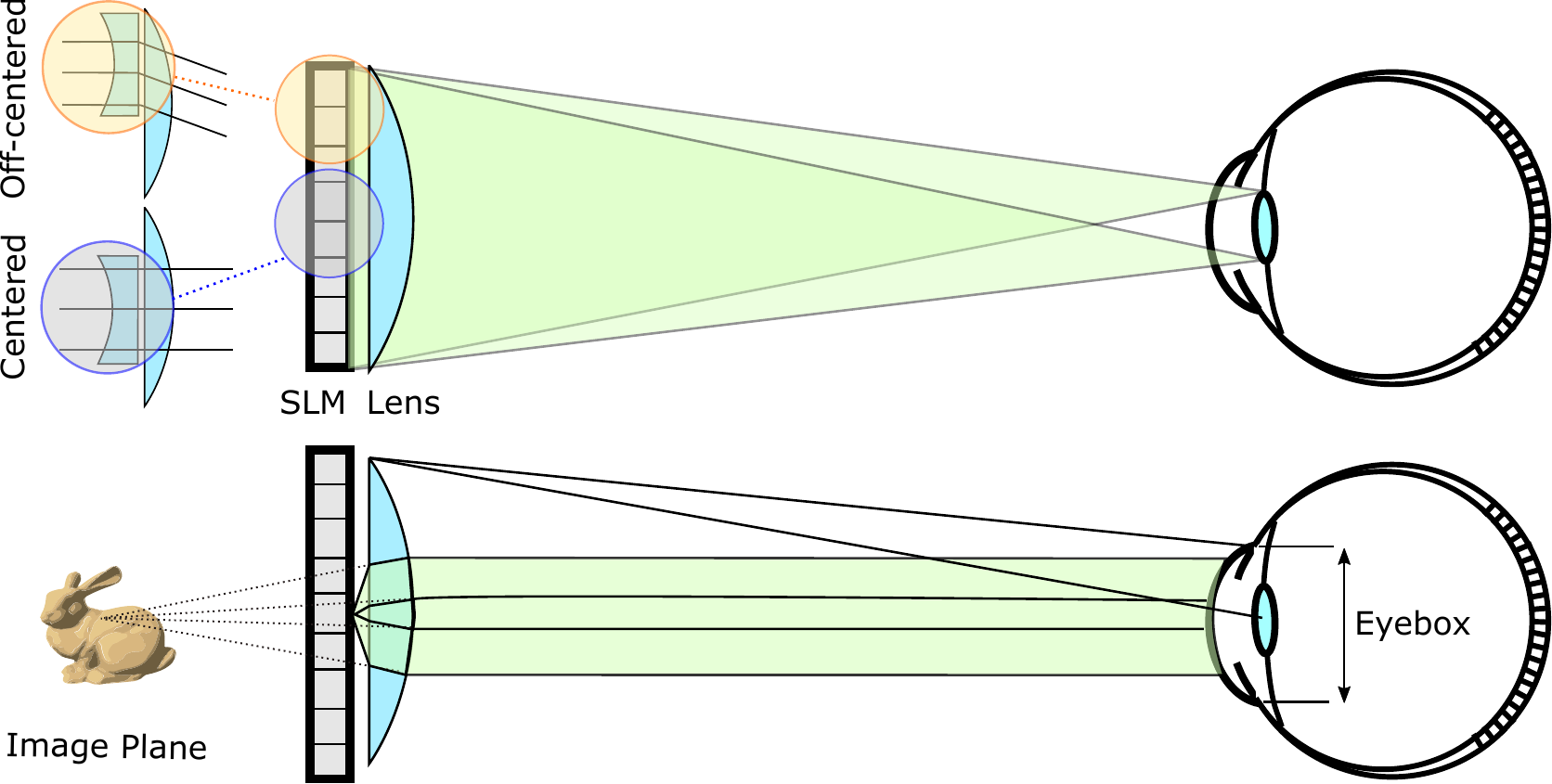}
    \caption{
    To evaluate the proposed pupil-aware holographic phase retrieval, we use a ray-biasing convex lens coplanar with the SLM, achieving a large eyebox. A conjugate lens phase function is applied on the SLM (top) which results in plane waves leaving the SLM-lens system (bottom), thereby creating a virtual image at optical infinity.}
    \label{fig:pupil_display}
\end{figure}

In this section, we introduce the proposed pupil-aware holographic phase retrieval algorithm. To formalize the image formation method and validate the proposed algorithm, we rely on a wide eye box holographic display as a testbed. To effectively use {\'etendue}, we use a single lens co-planar to the SLM to off-load a significant amount of SLM bandwidth (i.e., the supported spatial frequencies), see Fig.~\ref{fig:pupil_display}. While similar hardware configurations have been used in the past, where a lens is used as an eyepiece, to achieve a large FOV~\cite{maimone2017holographic}, we devise a variant for a larger eyebox which allows us to evaluate effects due to pupil variance. In the following, we first describe this setup. Then, we derive a pupil-aware image formation model for this setup, and optimize an SLM phase pattern for image quality and energy distribution throughout the eyebox.

\paragraph{Bandwidth-preserving Ray Biasing}
For a holographic display, {\'e}tendue (or space-bandwidth product) is the product of the area of the SLM and the maximum diffraction angle $\theta$ supported by the SLM. For a given wavelength of light $\lambda$, this angle is determined by the pixel pitch $p$ as $\theta = sin^{-1}(\lambda/2p)$. In other words, the supported field of view, without magnification optics, is only twice the diffraction angle, which is no more than $4^\circ$ for an SLM of $8\mu m$ pixel pitch, while the eyebox is the size of the SLM. To achieve a field of view of about $90^{\circ}$, the SLMs would need to have pixels of size less than $0.3\mu m$ -- today's SLM pixels are more than an order of magnitude larger. 

Under these limitations, unlike the existing display setups which maximize the FOV at the expense of the eyebox, we use a lens to bias rays towards the eyebox to get the desired ratio of a large eyebox and a limited FOV, given our available {\'e}tendue. Specifically, we quadratically offset phase with a thin lens, placed directly co-planar with the SLM as shown in Fig.~\ref{fig:pupil_display}. As a result, a plane wave incident on this SLM-lens system will bias incoming light and \emph{bend light towards the eyebox}. As the ray-biasing lens supports the FoV, the majority of the SLM bandwidth thereby is used to effectively increase the eyebox of the system. For example, off-loading the bandwidth of an SLM of pixel pitch $8\mu m$ to a lens of focal length $100mm$ achieves a large eyebox of size about $7mm$.

\begin{figure*}[t!]
     \centering
      \includegraphics[width=1.1\linewidth]{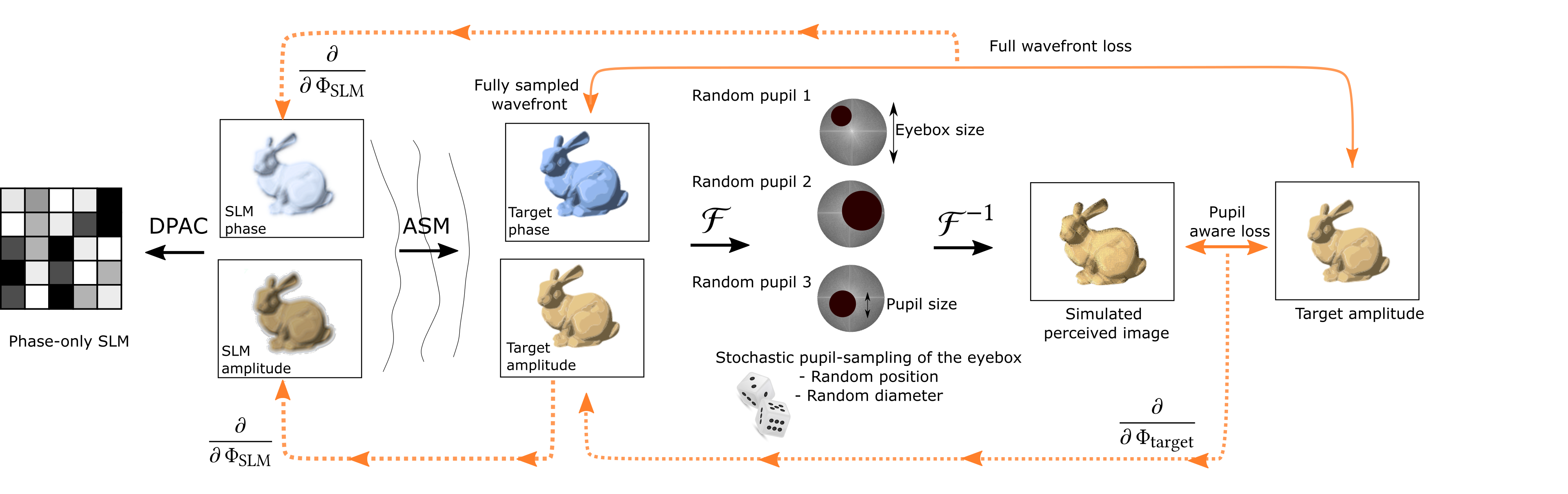}
      \vspace*{-20pt}
     \caption{%
     \emph{Pupil-aware Holography}. We optimize for phase-only holograms to produce high-fidelity reconstructions and energy distribution across the eyebox. To this end, we rely on a differentiable image formation model that explicitly considers the eye pupil sampling of the eyebox, allowing us to backpropagate an intensity reconstruction loss to the SLM phase pattern for each step of our iterative optimization. By a stochastically sampled pupil-aware optimization, the proposed method is able to achieve both image quality and energy distribution over the entire eyebox.}
     \label{fig:over}
\end{figure*}

\paragraph{Hologram Formation Model}
As described above, we use a convex lens coplanar with the SLM and generate a virtual image one focal length away from the lens (SLM plane) by algorithmically incorporating a (conjugate) concave lens kernel of the same focal length, as shown in Figure.~\ref{fig:pupil_display}. Note that this places the final image at optical infinity as the convex lens and the conjugate lens kernel effectively compensate each other. See Supplementary Material for additional details.
Specifically, for any given plane wave illumination, the conjugate lens phase on the SLM deflects the light in a diverging manner to form a virtual image. This diverging light is collected by the convex lens and bent to form a parallel beam. Like a physical diverging concave lens, the deflection of light near the center of the hologram is smaller whereas the deflection is larger at the edges. However, note that the SLM can only support a maximum deflection equal to that of the maximum diffraction angle $\theta$. Therefore, we confine the conjugate lens phase corresponding to a given target image pixel to this smaller diffraction support as illustrated in Figure.~\ref{fig:pupil_display}(top). A local support region for a given image pixel can be thought of as a \emph{central lens cut} from a larger concave lens. However, note that different \emph{lens cuts} corresponding to different object points can overlap on the hologram plane.

For a given target image field $U_\text{target} = Ae^{j\phi_0}$, where $A$ is the image amplitude and $\phi_0$ is the (unknown) object phase, a complete complex hologram $H$ on the SLM plane can be formed by computing the continuous sum of (overlapping) concave \emph{lens cut} phase functions weighted by the target pixel amplitudes. This complex hologram formation can be mathematically represented as
\begin{equation}
    H(\bar{x}) = \int_{\bar{t}\in L} A(\bar{x})e^{j\phi_0(\bar{x})} e^{-\frac{jk}{2f}(\bar{x}-\bar{t})} d\bar{t},
    \label{eq:img-formation-integral}
\end{equation}
where $\bar{x}$ is a target point and $\bar{t}$ is the offset translation of the concave lens as illustrated by the yellow inset in Figure.~\ref{fig:pupil_display}(top), and $L$ denotes the pixels within the diffraction angle support region that the SLM provides. Note that the above Eq.~\eqref{eq:img-formation-integral} is a convolution. Hence, we can define the image formation forward model as 
\begin{equation}
    H = U_\text{SLM} = U_\text{target} * G,
    \label{eq:img-formation-model}
\end{equation}
where $*$ is the convolution operator, $H$ is the complex hologram field on the SLM, $U_\text{target}$ is the target image field, and
\begin{equation}
    G = e^{-j\frac{k}{2f}(x^2+y^2)}
    \label{eq:kernel}
\end{equation}
is the field propagation kernel defined within the SLM diffraction angle support. Note that $f$ here is the focal length of the convex lens co-planar with the SLM. The convolution above can be computed in Fourier domain, thereby posing it as a band-limited angular spectrum propagation function~\cite{goodman2005introduction}.
Note that this image formation forward model \emph{creates a coherent planar wavefront traveling towards the eyebox}, which is located at the focal plane of the ray-biasing lens.

\paragraph{Learning Pupil-aware Holograms}

The object phase on the target image plane is responsible for the light distribution and hence energy distribution within the eyebox of the display. Note that although the eyebox supported by a display is larger, having the light energy concentrated in one region results in a smaller \emph{effective eyebox}, see Figure~\ref{fig:simulation_results2}. On the other hand, it also decides the quality of holographic images; a uniform object phase results in noise-free images with most energy concentrated into an extremely tiny effective eyebox whereas a completely random object phase results in extremely noisy images but with energy spread uniformly over the entire eyebox. As such, existing methods typically prefer a uniform phase on the target image plane for the quality of holographic projections. Although we can design a display that allows for a reasonably large eyebox, existing methods ignore that the \emph{effective eyebox} where all the light energy is concentrated may still be very small as shown in Figure~\ref{fig:simulation_results2}. As a result, the eye pupil movement outside of this \emph{effective eyebox} results in losing a significant portion of the image, and sometimes the image itself.

In order to tackle these problems, we jointly optimize for the SLM phase, holographic image quality and the eyebox energy distribution to support diverse eye pupil states within the eyebox. We derive a differentiable pupil-sampled variant of the forward model from the previous section to propagate the wave field effectively between the hologram plane, image plane and the eyebox, and stochastically sample the eyebox to incorporate different pupil states. Using this differentiable forward model, we learn the SLM phase using stochastic gradient descent with Wirtinger gradients~\cite{chakravarthula2019wirtinger,peng2020neural} in an iterative fashion. Integrated into the stochastic gradient descent scheme, we randomly sample different sizes and locations of the pupil within the eyebox, during the optimization. We initialize the optimizer with the hologram computed using the model discussed in the previous paragraph. 

Specifically, for a given target image amplitude $A_\text{target}$, we optimize for the complex hologram field at the SLM $U_\text{SLM} = A_\text{SLM}e^{j\Phi_{\text{SLM}}}$ to produce high-fidelity reconstructions as well as energy distribution across the eyebox, and consequently indirectly optimize the phase distribution of the propagated wavefront on the image plane. To this end, we model the \emph{reconstructed image at the image plane in two different ways.} We first propagate the incident wave from the SLM plane to the virtual image plane, filtering higher frequencies and orders within a 4F system, resulting in the fully-sampled wavefront simulated image on the target plane $U_\text{target; full}$ as

\begin{equation}
\label{eq:image_formation_forward}
U_\text{target; full} = \mathrm{Prop}_{\text{SLM} \rightarrow \text{target}}(U_\text{SLM}) = \mathcal{F}^{-1} \left( \mathcal{F}(U_\text{SLM}) \odot M_\text{iris} \right) * G^{\dagger},
\vspace{2mm}
\end{equation}

where $\mathrm{Prop}_{X \rightarrow Y}(U)$ is the wave propagation operator, propagating a field $U$ from the plane $X$ to $Y$, $M_\text{iris}$ is a binary mask used to filter higher orders within a 4F system and $G^{\dagger}$ is the complex conjugate of the kernel introduced in Eq.~\eqref{eq:kernel}. We compute the convolution in Eq.~\eqref{eq:image_formation_forward} in the Fourier space, specifically as a modified angular spectrum propagation between the SLM and the target image planes.

In addition, we also incorporate a forward model that includes the eye pupil. We propagate the complex wave field from the SLM to the eyebox, attenuate the eyebox with a stochastically sampled eye pupil mask $M$ and then propagate the sampled eyebox field to the image plane to reconstruct the image as seen by the eye pupil. Note that in our display configuration, the eyebox plane lies at the Fourier plane of the ray-biasing lens co-planar with the SLM. Hence the wave propagation between the target image plane and the eyebox plane can be modeled by the far-field Fraunhofer propagation. Therefore, the pupil-sampled reconstructed image $U_\text{target;pupil}$, can then be expressed as

\begin{equation}
U_\text{target;pupil} = \mathrm{ \mathcal{F}^{-1}}\left(M \odot \mathcal{F}(U_\text{SLM}) \odot M_\text{iris} \right) * G^{\dagger}.
\vspace{2mm}
\label{eq:image_formation_backward}
\end{equation}

The image formation between the SLM and target image planes ensures that the SLM phase produces the appropriate noise-free reconstructions when the entire wavefront is sampled. The image formation model between the target image plane and the stochastically sampled eyebox plane ensures that the image quality is maintained for a variety of pupil states across the eyebox. Therefore, we use both fully-sampled and pupil-sampled reconstructed images to find the optimal complex wave field at the SLM ($U_\text{SLM} = A_\text{SLM} e^{j\Phi_\text{SLM}}$) by solving the following optimization problem
\begin{equation}
A_\text{SLM}, \Phi_\text{SLM}  =\argmin {A'_\text{SLM}, \Phi'_\text{SLM}}  \mathcal{L}(|U_\text{target;full}|, A_\text{ref}) +      \mathcal{L}(|U_\text{target;pupil}|, A_\text{ref}),
\label{eq:loss-fun}
\end{equation}
where $A_\text{ref}$ is the desired reference image and $\mathcal{L}$ is a custom penalty function which we detail below. $U_\text{target;full}$ and $U_\text{target;pupil}$ are as defined in Eq.~\ref{eq:image_formation_forward} and Eq.~\ref{eq:image_formation_backward} respectively. Figure~\ref{fig:over} provides an illustration of this method. 

Within the proposed optimization, we employ stochastic sampling of eyebox to account for a variety of pupil states. This also allows for a \emph{pupil-aware} distribution of energy within the eyebox.
Optimizing for a complex SLM wave field allows for more degrees of freedom than a phase-only optimization, and offers an optically elegant way of removing noise from the images. For the noise removal, we filter higher frequencies corresponding to the complex wave field on the SLM during the optimization. This forces noise to the higher diffraction orders. To optically filter these higher diffraction orders carrying the noise, we encode the normalized complex SLM wave field (i.e., $Ae^{j\phi} \in \mathbb{C} (0\leq A \leq 1)$) into a phase-only pattern by interleaving the amplitude and phase using a double phase amplitude coding method~\cite{hsueh1978computer} 
\begin{equation}
    Ae^{j\phi} = 0.5e^{j(\phi-cos^{-1}A)} + 0.5e^{j(\phi+cos^{-1}A)} .
\end{equation}

We use a checkerboard mask to select the interleaving phase values from the decomposed two phase-only holograms. The interleaving into the high frequency checkerboard pattern result in high diffraction orders carrying high frequency noise which we filter in the Fourier plane.

We solve the above optimization problem using first-order iterative stochastic gradient descent methods as both the forward and backward image formations are differentiable with respect to the complex wave field at the SLM. 
For the image loss function $\mathcal{L}$, we use a weighted combination of $\ell_2$ penalty $\mathcal{L}_{\ell_2}$, SSIM $\mathcal{L}_\textsc{ssim}$, perceptual penalty based on VGG-19 $\mathcal{L}_\textsc{perc}$~\cite{Johnson2016PerceptualLF}, and Watson FFT $\mathcal{L}_\textsc{wfft}$~\cite{Czolbe2020ALF}, that is:
\begin{equation}
\mathcal{L} = \lambda_{\ell_2}\mathcal{L}_{\ell_2} + \lambda_\textsc{ssim}\mathcal{L}_\textsc{ssim} + \lambda_\textsc{perc}\mathcal{L}_\textsc{perc} + \lambda_\textsc{wfft}\mathcal{L}_\textsc{wfft}.
\end{equation}
We use a least-square penalty $\mathcal{L}_{\ell_2}$ for per-pixel accuracy in the reconstruction and $\mathcal{L}_\textsc{ssim}$ as a hand-crafted perceptual loss. The second perceptual penalty $\mathcal{L}_\textsc{perc}$ compares the image features from activation layers in a pre-trained VGG-19 neural network, that is,
\begin{equation}
\mathcal{L}_\textsc{perc} = \sum_{l} v_l || \phi_l(x) - \phi_l(y) ||_1,
\end{equation}
where $\phi_l$ is the output of the $l$-th layer of the pre-trained VGG-19 network and $v_l$ are the corresponding penalty-balancing weights. Specifically, we use the outputs of ReLU activations just before the first two maxpool layers, i.e., relu1$\_$2 and relu2$\_$2. The loss term $\mathcal{L}_\textsc{perc}$ therefore helps recover finer details. However, note that the VGG-19 network is optimized for classification and detection tasks, and is ``robust'' to the perceptual influence of artifacts such as noise. Therefore, we further augment the reconstruction quality by adopting the Watson FFT error function which is crafted specifically for human visual system, based on Watson's visual perception model~\cite{watson1993dct}, see Supplemental Document. We find the combination of the above losses helps steer the optimization towards holograms that maintain high quality across the eyebox.

\section{Implementation}
\label{sec:implementation}

\subsection{Software} 
We tested our pupil-aware holography concept in simulation implemented using PyTorch running on an NVIDIA P100 GPU. PyTorch now provides complex Wirtinger gradients within its auto-differentiation modules making implementation of the optimization scheme with state-of-the-art first order optimizers straightforward. We notice that different optimizers result in slightly different reconstruction quality. We use the Adam optimizer for solving Eq.~\eqref{eq:loss-fun} with a learning rate of $0.01$ and other default parameters. We assume an SLM of $1080 \times 1920$ pixel resolution with a pixel pitch of $8\mu m$ to match our physical hardware setup. All the input images and the holographic phase output was maintained at $1080 \times 1920$ resolution. The simulated Fourier space filter and the eye pupil were chosen to match the physical size of the prototype display. Specifically the Fourier space filter dictates the size of the eyebox in our display prototype and is maintained at $6mm$. We initialize the optimization with the non-pupil-sampling hologram modeled by Eq.~\eqref{eq:img-formation-model}, and run the optimizer for 500 iterations until convergence. The overall optimization takes about 2~seconds.

\begin{figure}[ht!]
    \centering
    \includegraphics[width=\linewidth]{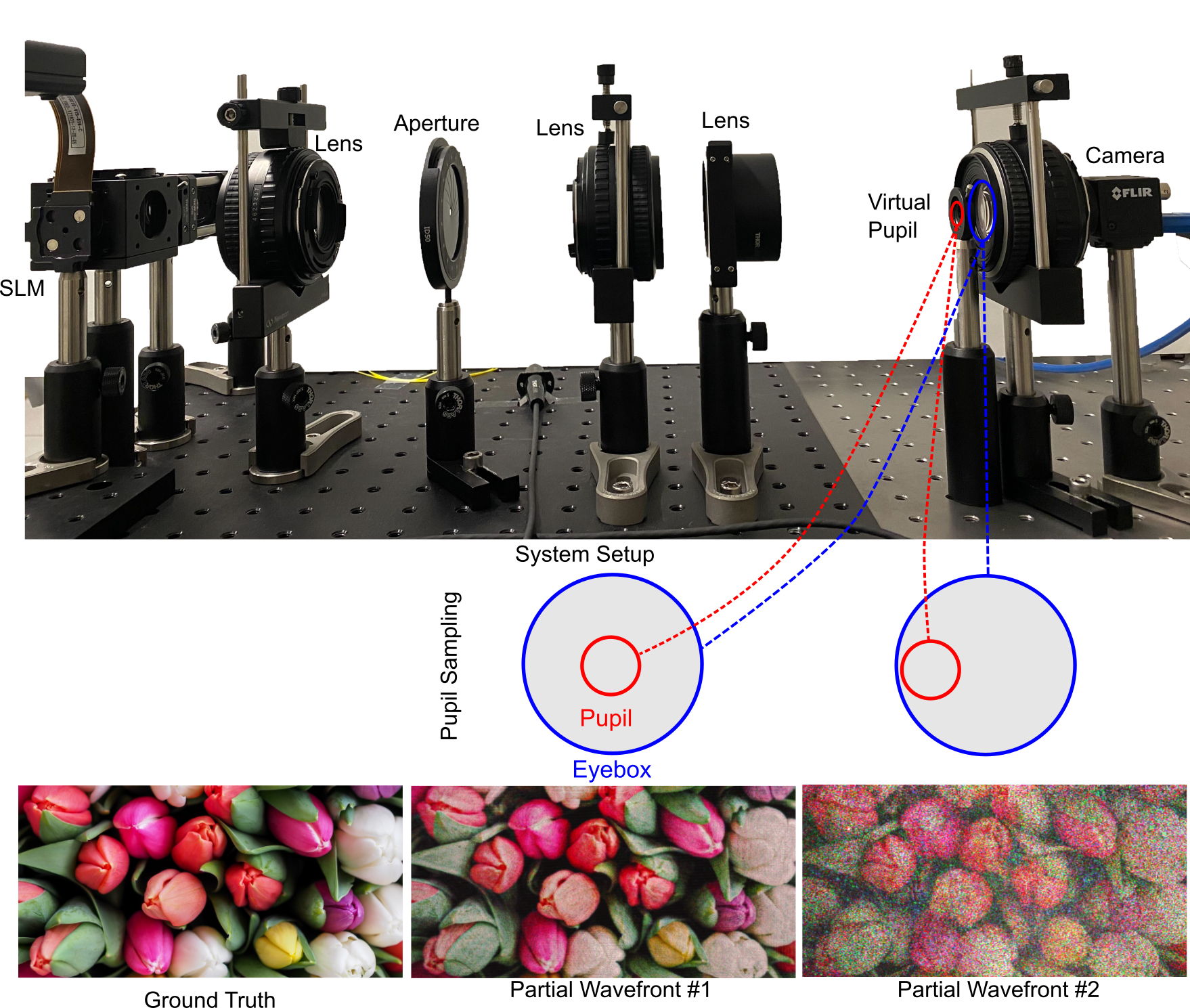}
    \caption{\emph{Prototype Holographic Display}. We built an experimental display to validate pupil-aware holographic phase retrieval. To mimic the pupil sampling of a human eye, we use an aperture on the eyebox plane, as can be seen in front of the camera, as a virtual pupil. The proposed pupil-aware holography method enables accurate holographic image reconstructions across the eyebox for diverse pupil states.
    }
    \label{fig:setup}
\end{figure}

\begin{figure*}[ht!]
    \centering
    \includegraphics[width=\linewidth]{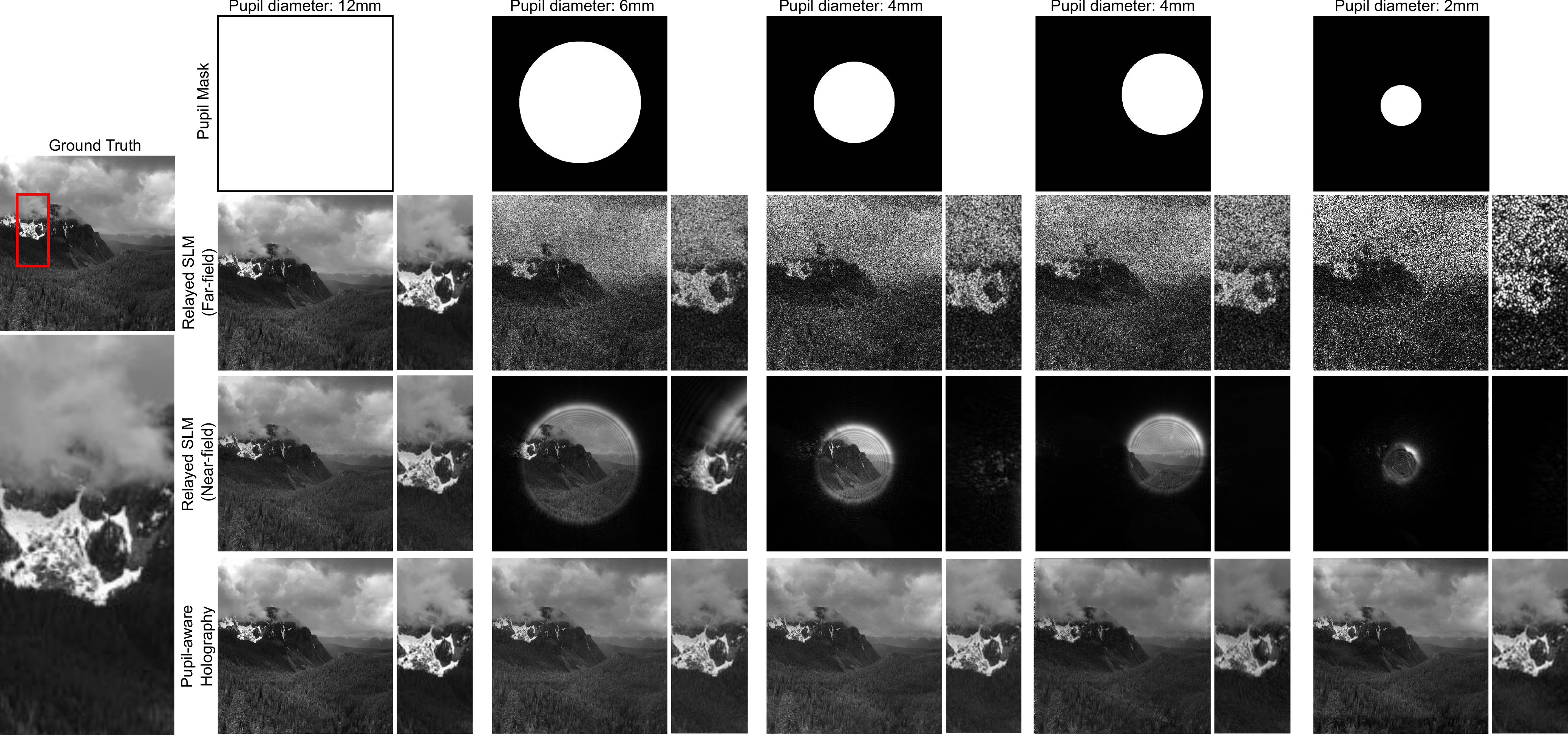}
    \caption{\emph{Evaluating Display Configurations (Simulation)}. 
    We study commonly used near- and far-field display configurations and the proposed wide eyebox variant for evaluating our pupil-aware holography method. 
    See Fig.~\ref{fig:baseline_display_designs} for SLM relayed and non-relayed setup schematics, and Fig.~\ref{fig:nonrelayed} for a small pupil forming Maxwellian-style display. 
    We test the configurations for five pupil masks with different sizes and locations shown on top, where the white mask represent the eye pupil. All existing methods, with varying pupil states, produce either speckle while truncating the spatial frequencies that are admitted into the pupil in the far-field configuration (second row), or truncate the image itself as shown in the near-field configuration (third row). The proposed variant of the wide eyebox display allows us to study pupil-aware holography that incorporates diverse pupil states .
    }
    \label{fig:simulation_results}
\end{figure*}

\subsection{Hardware Prototype} 
To validate our simulation, we built a prototype holographic display shown in Figure~\ref{fig:setup} with a HOLOEYE PLUTO liquid crystal on silicon (LCoS) reflective phase-only spatial light modulator with a resolution of $1920 \times 1080$ and a pixel pitch of $8\mu m$. This SLM is illuminated by a collimated and linearly
polarized beam in a color sequential manner from optical fibers emitting at a wavelengths of $636 nm$, $520nm$ and $450nm$ and controlled using a ThorLabs KLD101 Kinesis K-Cube laser diode driver.
The illuminated beam that is modulated by the phase-only SLM
is focused by a Pentax 645N 75mm  on an intermediate plane where an iris is placed to discard higher diffraction orders and conjugate images. We then relay the SLM with another Pentax 645N 75mm lens. Note that both the 75mm Pentax lenses form a 4F system with unit magnification, relaying the filtered SLM onto a virtual SLM plane as shown in Figure~\ref{fig:setup}. We then place a Thorlabs AC-508-100-A-ML $100mm$ focal length achromatic doublet coplanar with the virtual SLM to create ray-biasing as described in Section~\ref{sec:pupil-invariant}. This configuration created an eyebox of $6mm$ where we place an iris on a motorized translation stage for sampling the eyebox. The sampled eyebox field is then measured by a Point Grey FLIR machine vision camera with a focusing lens. 

We also built prototype displays for the relayed-SLM configuration showing both near-field and far-field holograms. As comparable holographic setups have been described in detail in a large body of existing work, we refer to the Supplemental Document for more details on the hardware configuration of these setups.

\section{Assessment and analysis}
\label{sec:assessment}

\subsection{Synthetic Evaluation}

\label{subsec:synthetic_eval}
Next, we investigate the pupil dependence of image quality in simulation for different near- and far-field display configurations (see Figure~\ref{fig:baseline_display_designs}) as described in Section~\ref{sec:fwdmodels}. We then evaluate the effectiveness of pupil-aware holographic phase retrieval on the proposed wide eyebox display configuration (see Figure~\ref{fig:pupil_display}) and compare it to non-pupil-aware holographic reconstructions on the \emph{same} display. For these simulations, all the input images and output holograms used a resolution of $1920\times 1080$ and an SLM pixel pitch of $8 \mu m$ to match our hardware. Finally, we discuss extending our pupil-aware holography method to larger {\'e}tendue displays, validating it on $16\times$ and $64\times$ larger {\'e}tendue.

\paragraph{Evaluating Different Display Configurations}
Figure~\ref{fig:simulation_results} compares the fidelity of the reconstructed holograms for three different display configurations in diverse pupil states.
Specifically, we show five different pupil states including the 12\,mm virtual pupil diameter that samples the entire wavefront. While humans have pupils that span a diameter of around 4~mm in normal conditions, we note that a very large synthetic pupil diameter of 12\,mm is a commonly used setting in various conventional holographic displays~\cite{chakravarthula2020learned,peng2020neural}.
In this unrealistically large pupil state, the entire wavefront impinging the pupil is collected by the eye lens and focused to the retina. Another way to achieve this is to use an eyepiece to relay the wavefront into the eye pupil, but this reduces the size of the eyebox to typically less than a millimeter, and hence the image can completely disappear even for minute eye pupil movements, see Figure~\ref{fig:nonrelayed}.
Collecting the full wavefront ensures high-fidelity reconstruction of holographic images across all tested configurations as shown in the first column of Figure~\ref{fig:simulation_results}.
We next evaluate the configurations on more realistic pupil parameters in the range from 2\,mm to 6\,mm in different locations.
This results in partial sampling of the modulated wavefront as shown in the pupil masks at the top row of Figure~\ref{fig:simulation_results}.
For the far-field relayed SLM configuration, the pupil diameter mainly determines the degree of artifacts such as the speckle noise, while the location of the pupil does not affect the overall performance. This is due to the random phase distribution typical to the far field holograms which results in a relatively uniform but random energy distributed within the eyebox. 
In contrast, pupil sampling acts as a cropping operation for near-field holograms with additional possible diffraction artifacts due to the size of the pupil and the propagation distance of the projected light. 
The qualitative difference between the far-field and the near-field setups manifests in whether the image degradation appears globally or locally on the retina plane. In particular, displaying a near-field hologram, which is generally preferred due to their high-fidelity reconstructions relative to the far-field, mandates that \emph{the eyebox is always smaller than the size of the eye pupil} for high-fidelity reconstructions. The proposed method provides accurate holographic reconstructions across diverse pupil states including both pupil diameter and locations as shown in Figure~\ref{fig:simulation_results}. This is because we model the image formation at the SLM plane as a near-field propagation thus ensuring high-fidelity reconstructions, the final image received by the eye is that of a far field due to the lens placed co-planar with the SLM thus ensuring a sufficiently large effective eyebox.

\begin{figure*}[t!]
    \centering
    \includegraphics[width=\linewidth]{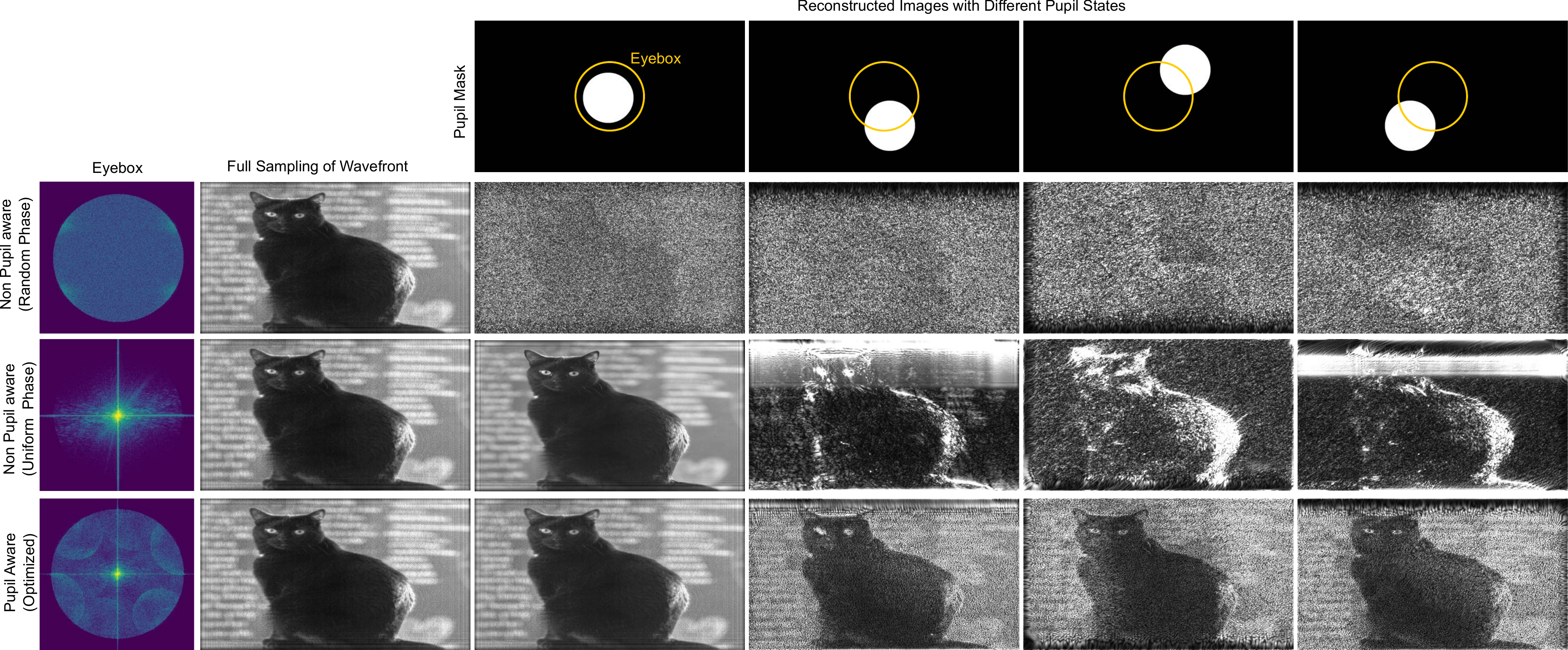}
    \caption{\emph{Pupil-aware vs non-pupil-aware holography on the same display setup (Simulation)}. Conventional non-pupil-aware holographic phase retrieval assumes an unknown object phase, which is typically either random (top row) or uniform zero (middle row). While a random object phase on the image plane generates a uniform eyebox energy, it results in the reconstructed image corrupted with extreme noise (top row). On the other hand, a uniform object phase results in energy concentrated within the eyebox, which results in noise-free reconstructions at the center of eyebox and loss of image as the pupil moves away from it, resulting in a tiny effective eyebox (middle row).
    The proposed pupil-aware holography results in high-fidelity reconstructions by distributing the energy across the eyebox (bottom row).
    }
    \label{fig:simulation_results2}
\end{figure*}

\paragraph{Evaluating Pupil-aware Phase Retrieval on the Same Display}
We next analyze the effectiveness of the proposed pupil-aware holography optimization on the proposed display configuration, compared to conventional phase retrieval methods where the unknown object phase is chosen to be either random or uniform~\cite{maimone2017holographic,shi2021towards}. 
Note that the choice of the object phase on the target image plane effects both the reconstruction quality of holographic projections as well as the energy distribution within the eyebox. 
Figure~\ref{fig:simulation_results2} shows the simulated energy distribution at the eyebox plane for three different experiments on the same display configuration. The first column shows the eyebox energy distribution for random and uniform object phase, and for pupil-aware optimized eyebox. 
A completely random object phase results in a uniform distribution of light in the eyebox, but at the cost of image quality for different pupil states, see Figure~\ref{fig:simulation_results2} (top row).
On the other hand, although the size of the eyebox supported by the display is larger, a hologram computed using a uniform phase on the target plane results in energy concentrated in the center, see Figure~\ref{fig:simulation_results2} (middle row). Therefore, an image is visible only if the eye pupil samples the center of the eyebox where the maximum intensity is found. Any deviation from the center results in a loss of image. This reduces the eyebox effectively to a very small region in the center. However, with our pupil-aware optimization, we notice that the energy in the eyebox is distributed similar to that of a constant phase in the center and approaches pseudo random phase towards the edge of the eyebox (Figure~\ref{fig:simulation_results2} (bottom row)), thus finding a tradeoff between both extremes. This energy distribution leads to higher fidelity reconstructions across diverse pupil states.

\paragraph{Analysis of Optimized Object Phase}

\begin{figure}
    \centering
    \includegraphics[width=\linewidth]{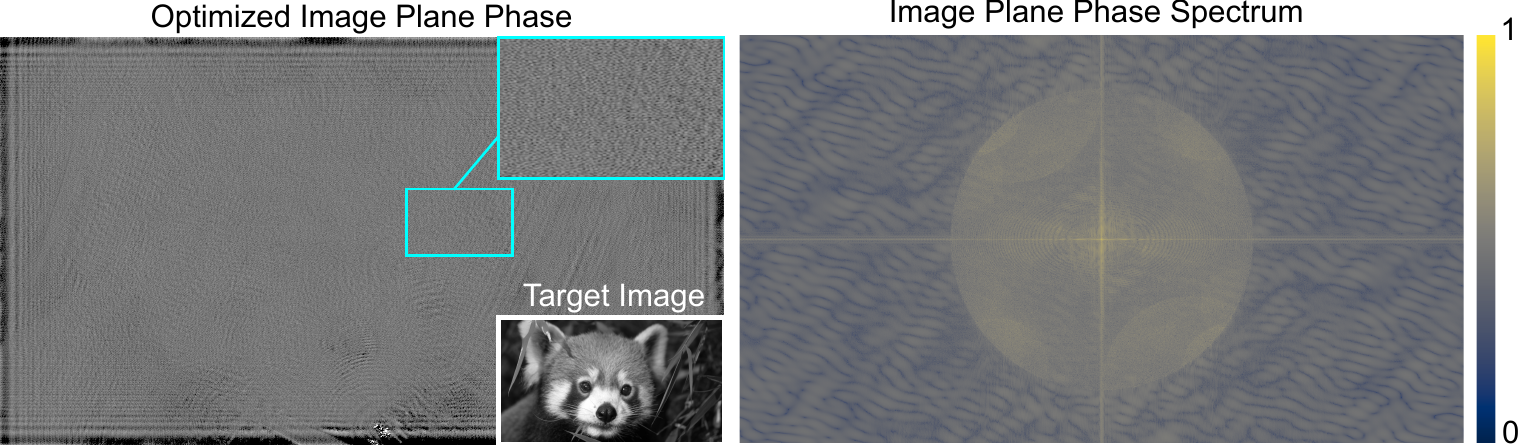}
    \caption{\emph{Optimized Object Phase and its Spectrum}. The optimized object phase resulting from the proposed method contains some of the target image structure overlaid with high frequencies. A log spectrum of this phase image shows the presence of low frequencies (similar to constant object phase) and high frequencies (similar to random object phase). The circular shape is from Fourier filtering of higher orders within a 4F system, which is also the size of the eyebox. This object phase results in a pupil-aware eyebox energy distribution.}
    \label{fig:target_phs}
\end{figure}

As discussed in Section~\ref{sec:pupil-invariant}, 
the object phase on the image plane effects the energy distribution on the eyebox plane and hence the effective size of the eyebox. On the other hand, it also effects the reconstructed image quality. 
A random object phase results in energy distributed across the eyebox but noisy images. A uniform object phase results in visibly less image artifacts but a small effective eyebox. 
Contrary to this, holograms learned using the proposed method result in high image quality as well as energy distributed across the eyebox. To understand this, we study the optimized target phase on the image plane as well as its frequency spectrum. An example is presented in Fig.~\ref{fig:target_phs}. We notice that the optimized object phase resulting from the proposed method perhaps surprisingly contains some of the structure of the image amplitude. We observe an impulse response in the center resembling that of uniform object phase. Most high frequency components of the target phase are to be limited to a disc of the size of the filter we employ in the Fourier plane to filter any higher orders. Repeating circular patterns appear within this region which we hypothesize results in meaningful images when the eye pupil diverges significantly from the eyebox. A similar distribution can be observed in the eyebox as well, see Fig.~\ref{fig:simulation_results2}. The higher frequencies outside of the central disc are filtered in the Fourier plane and hence might be carrying high frequency noise that is eventually filtered in the Fourier plane (optically with an iris in experimental hardware display).

\begin{figure}
    \centering
    \includegraphics[width=\linewidth]{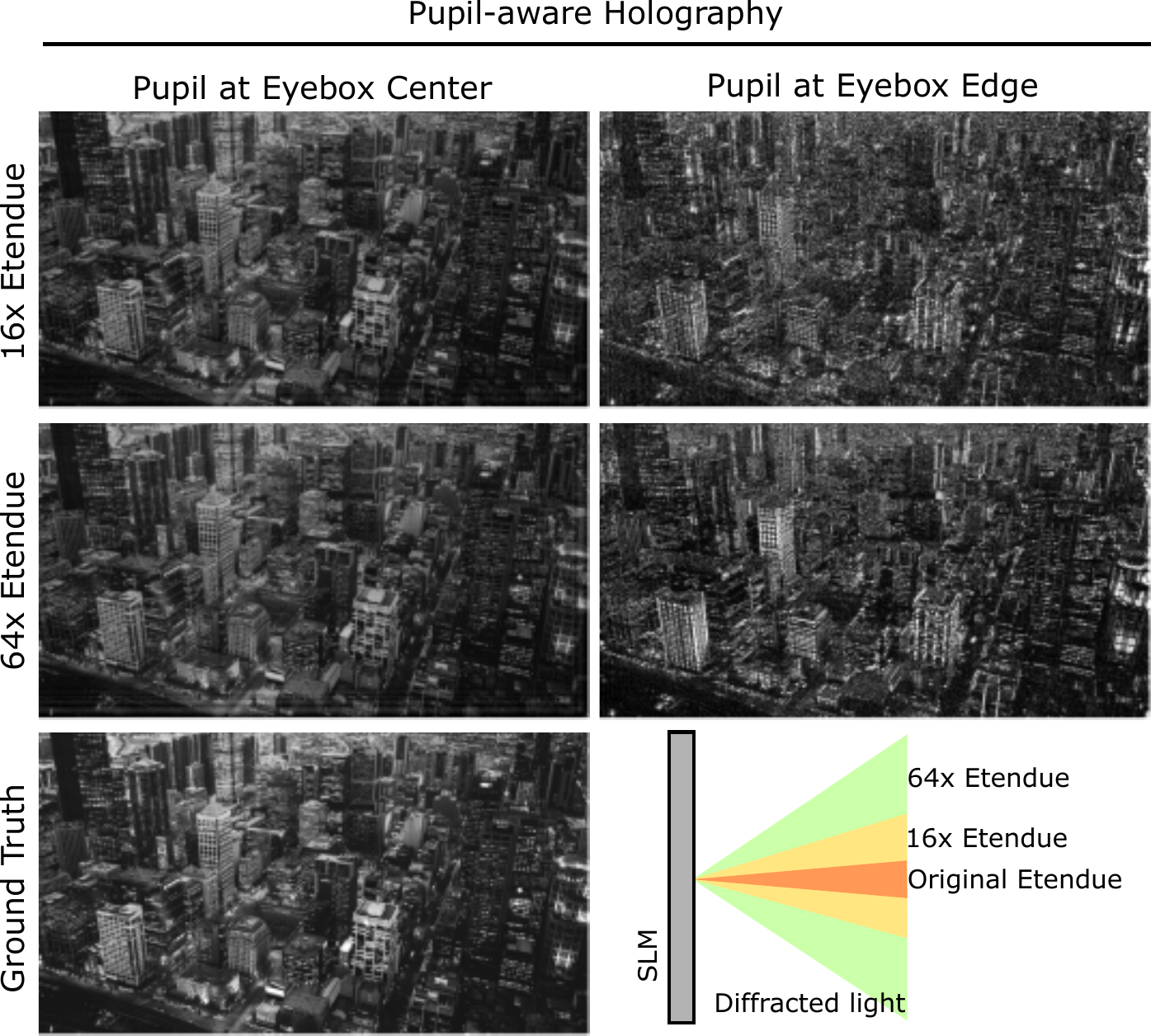}
    \caption{\emph{Pupil-aware Holography on Large {\'E}tendue Displays (Simulation)}. We apply pupil-aware holographic phase retrieval on $16\times$ and $64\times$ large {\'e}tendue display compared to our prototype hardware display. The results here demonstrate that our method scales to future large {\'e}tendue displays}
    \label{fig:etendue}
\end{figure}

\paragraph{Pupil-aware Holography on Large {\'E}tendue Displays}
Today's SLM pixel technology does not support large {\'e}tendue displays natively. Hence, we validate that pupil-aware holography expands to such future large {\'e}tendue displays in simulation. The simulations in the previous paragraphs match our hardware which uses an SLM of $1920\times 1080$ pixel count and a pitch of $8\mu m$. Next, we simulate a $16\times$ larger {\'e}tendue display with a pixel count of $7680 \times 4320$ and pixel pitch of $2\mu m$, and a $64\times$ larger {\'e}tendue display with a pixel count of $15360 \times 8640$ and a pixel pitch of $1\mu m$. The phase-only hologram SLM patterns are computed using the proposed method and report results in Figure~\ref{fig:etendue}. The results from smaller {\'e}tendue displays transfer to this setting and we validate that optimizing holograms in the pupil-aware fashion results in high-fidelity reconstructions. The proposed method scales well to larger {\'e}tendue displays as it only considers the pupil sampling and the energy distribution within the eyebox, independent of FoV of the display. We refer to the Supplementary Material for additional discussion on {\'e}tendue and the eyebox considerations. 

\subsection{Experimental Evaluation}

We validate the proposed pupil-aware holography method on a prototype display and report experimental findings.
Specifically, we first compare different display configurations with our wide eyebox display configuration (also assessed in simulation in the previous section) to validate the choice of our display setup. We then validate our pupil-aware holograms by comparing to non-pupil-aware holograms on the \emph{same experimental wide eyebox display configuration}. 

\paragraph{Validating the Wide Eyebox Display Configuration} 

\begin{figure*}
    \centering
    \includegraphics[width=\linewidth]{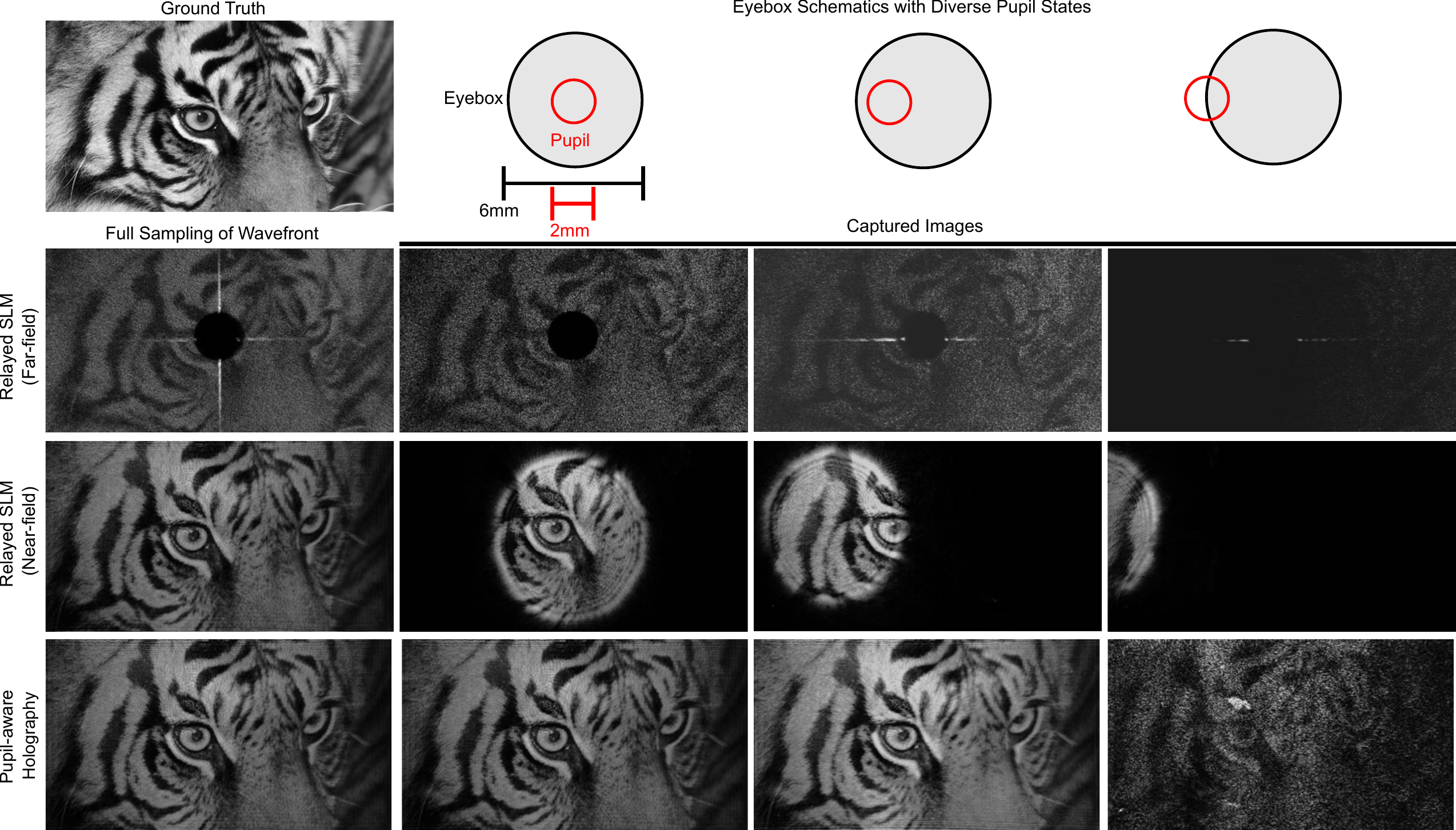}
    \caption{\emph{Experimental Evaluation of Various Display Configurations}. We validate the choice of our wide eyebox variant of holographic display compared to common existing display configurations by capturing the reconstructed holograms on a hardware prototype at three different pupil states shown in the top row. While the far- and near-field relayed SLM configurations allow for accurate reconstructions when the wavefront is fully sampled, they suffer from speckle and cropping artifacts due to partial pupil sampling (second and third row). In contrast, the proposed display configuration allows for evaluating pupil-aware holography for diverse pupil states across the eyebox. All holograms are computed using stochastic gradient descent iterative optimization.}
    \label{fig:real_results}
\end{figure*}

We next validate the proposed pupil-aware display configuration in conjunction with the proposes phase retrieval method. To this end, we display the SLM phase patterns for a given target image and measure the raw images with a 2\,mm-diameter pupil (iris) sampling the 7mm eyebox at different locations as shown in the top row of Figure~\ref{fig:real_results}. The top row of the figure indicates the pupil sampling of the eyebox. The subsequent rows show the measured images for far- and near-field configurations where the SLM is relayed onto the pupil, and our wide eyebox display configuration. Please see Supplementaty Material for more details on the display configurations. 

In the first column, we show the measured images for the three different configurations where the entire wavefront is sampled. We see that sampling the entire wavefront from the SLM results in high fidelity reconstructions. Note that measured images from the relayed SLM with far-field holograms have a dark region in the center due to a mask we used to block the high DC intensity of the unmodulated light from the SLM, see Supplementary Material for more details.
When the entire wavefront is sampled, the pupil-relayed near-field configuration (third row) and our wide eyebox configuration (fourth row) demonstrate comparable reconstructions. 
However, matching the simulations, partial sampling of the wavefront due to the pupil causes severe degradation in image quality in both near- and far-field relayed-SLM configurations (second and third row). Specifically, while the far-field holograms suffer from reduced brightness and increased speckle artifacts, the near-field holograms suffer from the image being cropped and diffraction artifacts at the edges of the pupil.
As the pupil moves towards the edge of the eyebox, these artifacts increase further until the entire image is lost in either cases, see last column of Figure~\ref{fig:real_results}. 
The proposed wide eyebox display configuration with pupil-aware holographic phase retrieval providing accurate reconstructions at different pupil states within the eyebox, as shown in the last row of Figure~\ref{fig:real_results}. Note that, at the edge of the eyebox, as opposed to losing the image entirely, pupil-aware holograms demonstrate reconstructions that look similar to the far-field holograms in relayed-SLM configuration, see last row of fourth column and second row of Figure~\ref{fig:real_results}. As discussed in Section~\ref{subsec:synthetic_eval}, the center of the eyebox of our optimized holograms achieves an energy distribution similar to that of uniform phase holograms resulting in noise-free reconstructions within the eyebox. As far eccentricities of the eyebox approach the distribution similar to a (pseudo) random phase, the reconstructions look akin to a random object phase hologram but still with visible image features, even when the pupil only partially overlaps with the eyebox. We do not evaluate the small eyebox Maxwellian-style display configurations~\cite{maimone2017holographic} here as the eyebox is typically less than a millimeter, as discussed in Section~\ref{sec:fwdmodels}.    
Overall, we find that our wide eyebox display configuration is best suited for evaluating the pupil-aware holographic phase retrieval method.

\begin{figure*}[ht!]
    \centering
    \includegraphics[width=\linewidth]{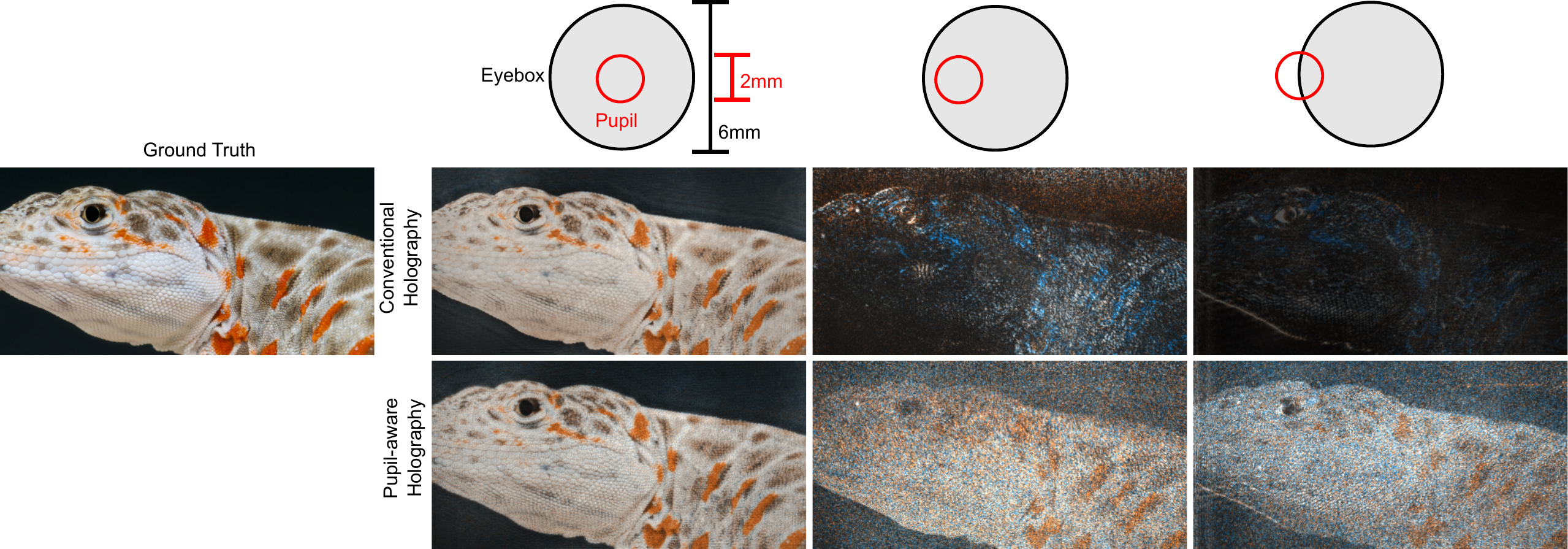}
    \caption{\emph{Experimental Evaluation of Pupil-aware Holography}. We validate the proposed pupil-aware holography on a wide eyebox hardware display setup, with the eyebox sampled by different pupil states (top). In contrast to conventional holographic methods that do not consider pupil sampling (middle), learning pupil-aware holograms lead to higher-fidelity reconstructions across the eyebox (bottom) only from partial wavefront sampling. }
    \label{fig:real_results_wo_w_pupil}
\end{figure*}

\paragraph{Validating Pupil-aware Holography}
We now validate the proposed pupil-aware holographic phase retrieval on the same wide eyebox display configuration described in Section~\ref{sec:pupil-invariant} and evaluated above. 
To this end, we display phase patterns computed with and without pupil-awareness, using the proposed optimization method in both cases (with and without pupil sampling), on an experimental prototype display and report results in Figure~\ref{fig:real_results_wo_w_pupil}. We also report a dense set of measured images with finer pupil sampling in Figure~\ref{fig:real_results_dense}. See Supplementary Material for additional results.

Figure~\ref{fig:real_results_wo_w_pupil} demonstrates the image fidelity achieved by our pupil-aware holography method over the eyebox. As described in the synthetic evaluations in Section~\ref{subsec:synthetic_eval}, the energy distribution achieved within the eyebox of our pupil-aware holograms approach that of noise-free uniform target phase holograms in the center (where the eye pupil is expected to sample the most) and that of a pseudo-random target phase holograms at higher eccentricities of the eyebox. The relative pupil position within the eyebox is indicated on the top. Holograms computed for maximum image quality but without pupil-awareness result in high-fidelity reconstructions when the pupil samples the center of the eyebox where maximum energy is concentrated. However, as the pupil moves away from the eyebox, the image is lost, see top row of Figure~\ref{fig:real_results_wo_w_pupil}. In contrast, our pupil-aware holographic phase retrieval distributes the energy throughout the eyebox to maintain the image quality as can be seen in bottom row of Figure~\ref{fig:real_results_wo_w_pupil}. As the pupil is likely to sample the center of the eyebox the most, majority of the SLM bandwidth is used to redirect the light to produce noise-free reconstructions in lower eccentricities of the eyebox. However, at higher eccentricities approaching the edge of the eyebox, image reconstructions that approach random phase holograms are produced. This is in accordance with the eyebox and object phase spectrum analysis reported in Section~\ref{subsec:synthetic_eval}. To better visualize the degradation of image quality, we report a finer pupil sampling of the eyebox and the corresponding holographic reconstructions in Figure~\ref{fig:real_results_dense}. It can be observed that the image quality of pupil-aware holograms is maintained at eccentricities where the images of non-pupil-aware holograms are completely lost. See Supplementary Material for additional results.

\begin{figure}
    \centering
    \includegraphics[width=0.89\linewidth]{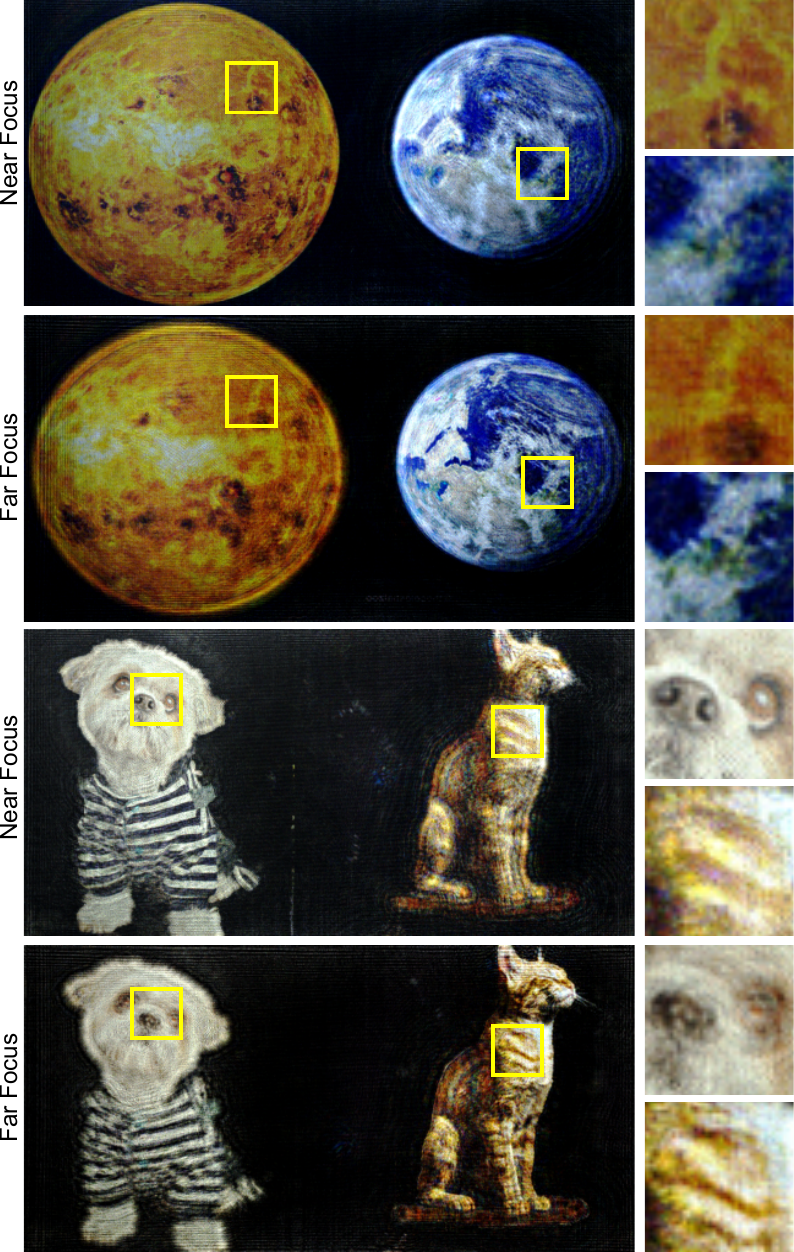}
    \caption{\emph{Experimental 3D Holographic Display}. 
    We demonstrate that the proposed method can also be extended to 3D holography. To this end, we show multiplane 3D holographic projections on an experimental propototype. That is, using the proposed method we optimize single SLM pattern to simultaneously project imagery at both near and far distances. The images shown here are measured only by changing the camera focus, and the corresponding in-focus and out-of-focus imagery can be observed in the insets.
 }
    \label{fig:real_results_3d}
\end{figure}

\paragraph{3D Pupil-Aware Holography}
Although studying eyebox dependence of holographic image quality is an unexplored area independently if 2D or 3D holograms are considered, the proposed pupil-aware holography also extends to 3D holographic displays and future very large {\'e}tendue displays. To validate extension to 3D holographic displays, we demonstrate a multiplane display application in Figure~\ref{fig:real_results_3d} where the holograms are computed using our pupil-aware holographic phase retrieval. The focus is changed between near and far objects by only changing the camera focus ring. For example, in the first column, the features on the dog such as the nose and the stripes on the shirt are sharply visible when the camera is focused to near distance whereas the stripes on the skin of the cat are blurred. When camera is focused at far distance, the cat comes to focus whereas the dog appears blurred. Similar trend can be observed in the second column of Figure~\ref{fig:real_results_3d} where the features on the Earth and Venus go in and out of focus at near and far distances. Some of these features are zoomed and shown in the insets for accessibility.

\begin{figure*}[ht!]
    \centering
    \includegraphics[width=0.98\linewidth]{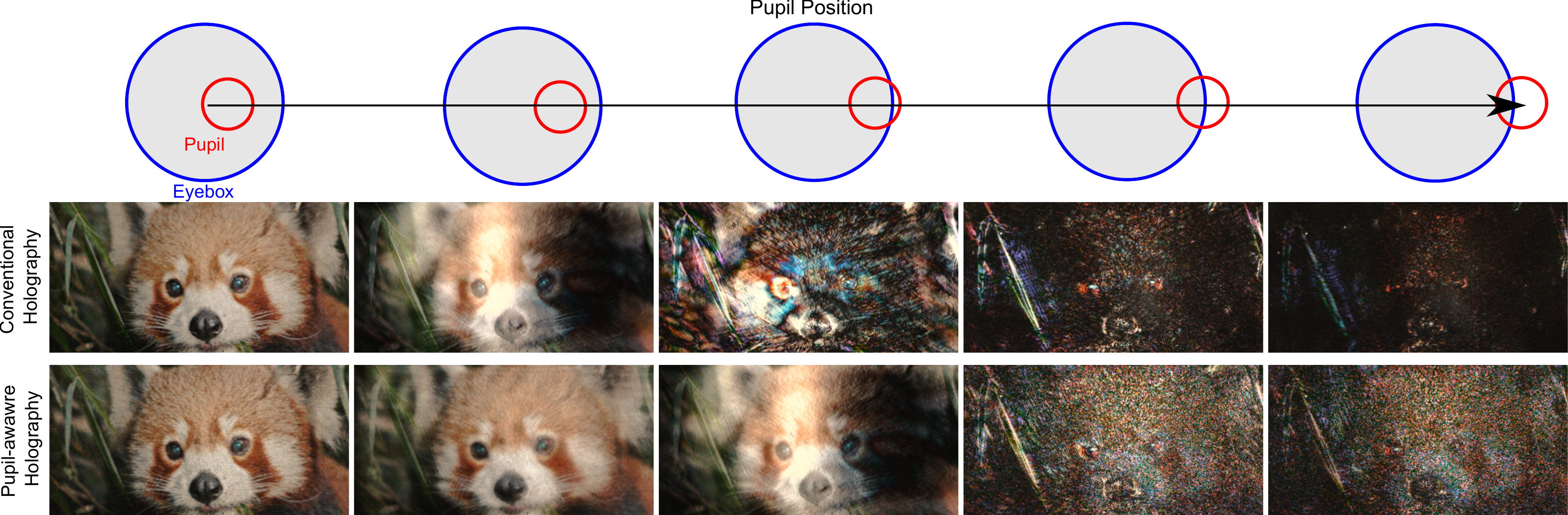}
    \caption{\emph{Experimental Evaluation of Densely-sampled Pupil States}. We densely sample the eyebox (top) to study the image quality as the pupil traverses the eyebox. Conventional holographic methods produce high-fidelity imagery at the center of the eyebox by concentrating the light energy distribution at the center. As a result, conventional non-pupil-aware methods results in complete loss of imagery as the pupil moves from the center (middle). Pupil-aware holography maintains the imagery throughout the eyebox as shown in the bottom row. }
    \label{fig:real_results_dense}
\end{figure*}

\section{Conclusion}
\label{sec:conclusion}
We propose a holographic display method that is pupil-aware. The rapid loss of image quality due to pupil sampling of the eyebox by the human eye has not been investigated in recent holographic approaches because the {\'etendue} that today's displays are restricted by is too small to observe this issue. We find that large {\'etendue} comes at the cost of a high degree of speckle, or drastic loss of overall intensity, and, as such, even with an ideal 1 billion pixel SLM, existing holographic displays are impractical as they are all fundamentally subject to pupil sampling of the wavefront at the eyebox. We investigate and explain such pupil effects when using only part of the eyebox. To generate pupil-aware holographic projections, we propose a content-driven display algorithm that generates high quality images across the eyebox regardless of the eye pupil size and location. We rely on a differentiable pupil-aware image formation model and a corresponding per-image optimization method that incorporate sampling over diverse pupil states, and, together, ensure image fidelity and energy distribution across the eyebox when solving for SLM patterns. We validate this method in simulation and with an experimental prototype system, where we achieve high image fidelity over the full eyebox extent. As such, we make a first step towards pupil-invariant large {\'etendue} displays of the future, which may make holography practical across application domains. Immediate next steps in this direction could be the joint optimization of custom {\'etendue}-expanding elements with pupil invariance -- potentially paving the way towards fully pupil-invariant holography in the future.


\bibliographystyle{ACM-Reference-Format}
\bibliography{references}
\end{document}